\documentclass[]{emulateapj}
\begin{document}
\title{An Improved Bandstrength Index for the CH G band of Globular Cluster Giants}
 
\bigskip
\author{Sarah L. Martell and Graeme H. Smith} 
\affil{University of California Observatories/Lick Observatory}
\affil{Department of Astronomy \& Astrophysics, UC Santa Cruz}
\affil{1156 High St, Santa Cruz, CA 95064}
\email{martell@ucolick.org, graeme@ucolick.org}
\and
\author{Michael M. Briley}
\affil{Department of Physics and Astronomy, University of Wisconsin Oshkosh}
\affil{800 Algoma Blvd, Oshkosh, WI 54901}
\email{mike@maxwell.phys.uwosh.edu}
\bigskip
\begin{abstract} 
\noindent

Spectral indices are useful tools for quantifying the strengths of features in moderate-resolution spectra and relating them to intrinsic stellar parameters. This paper focuses on the 4300$\mbox{\AA}$ CH G-band, a classic example of a feature interpreted through use of spectral indices. G-band index definitions, as applied to globular clusters of different metallicity, abound in the literature,  and transformations between the various systems, or comparisons between different authors' work, are difficult and not always useful. We present a method for formulating an optimized G-band index, using a large grid of synthetic spectra. To make our new index a reliable measure of carbon abundance, we minimize its dependence on [N/Fe] and simultaneously maximize its sensitivity to [C/Fe]. We present a definition for the new index $S_{2}(CH)$, along with estimates of the errors inherent in using it for [C/Fe] determination, and conclude that it is valid for use with spectra of bright globular cluster red giants over a large range in [Fe/H], [C/Fe], and [N/Fe]. 
\end{abstract}


\section{Introduction}
Spectroscopic indices are a measure of the strength of an absorption or emission feature, usually reported as the magnitude difference between the integrated flux in the region of the feature in question (the ''science band'') and one or two nearby continuum regions (the ''comparison bands''). Indices and narrow-band photometry have a long history of usefulness in terms of physical information returned per time spent observing. They can be used to study a wide range of stellar properties: Ca II H\&K surveys quickly identify extremely metal-poor stars (e.g., Beers et al 1985, 1999\nocite{B85}\nocite{B99}), the Mg I index is an indicator of surface gravity in cool stars (e.g., Morrison et al. 2000\nocite{M00}), and Balmer indices are diagnostics of stellar age and metallicity when seen in galaxy spectra (e.g., Kauffmann et al. 2003\nocite{K03}). Because they require far less observing time than high-resolution spectroscopy, the telescope-limited astronomer frequently uses these low-resolution methods in a first pass to identify interesting objects to be followed up later in more detail. 

Narrow-band colors are often used to calculate physical quantities: an empirical calibration can be established between the observed quantities and the intrinsic properties in a well-studied sample, and that calibration can then be applied to derive the intrinsic properties of a much larger set of stars, as long as the stars in the larger set are similar enough in temperature, gravity, and composition to the well-studied calibration sample. The \citet{A99} photometric temperature calibration is a well-known example of this process. The observables can also be used directly as proxies; for example, DDO and Str\"{o}mgren colors, which are known to correlate with temperature, metallicity, and surface gravity, can be used for dwarf/giant separation, or to do rough sorting by metallicity (e.g., Grundahl et al. 2002\nocite{GBN02}). The so-called Lick indices can be measured from integrated spectra of galaxies, and provide information about unresolved stellar populations (e.g., Worthey 2004\nocite{W04}).

Molecular bandstrength indices measured from low- or moderate-resolution spectroscopy are often the end result themselves: comparing CN versus CH bandstrengths of globular cluster red giants, one finds that CH bandstrength falls with rising luminosity \citep{C82} (the same is true in the halo field: see Gratton et al. 2000\nocite{GSCB00}), while at a given luminosity the CH-weaker (and therefore lower carbon) stars tend to have the stronger CN bands, and therefore must be nitrogen-enhanced. These trends imply that red giant atmospheres must be experiencing ongoing CN(O)-cycle processing (e.g., McClure \& Norris 1977 \nocite{MN77}, Shetrone et al. 1999\nocite{S99}). Much research into the behavior of C and N elemental abundances in globular cluster red giants (e.g., Briley \& Cohen 2001\nocite{BC01}, Norris et al. 1984\nocite{N84}, Norris \& Zinn 1977\nocite{NZ77}, Zinn 1973\nocite{Z73}) has relied heavily on the use of the 4300 $\mbox{\AA}$ CH bandstrength for [C/Fe] determination. This is true of Population II field giants as well (e.g., Rossi et al. 2005\nocite{R05}). However, there has been little uniformity in the CH indices used by various researchers.
Some G-band indices that have been used to date are $S(CH)$ \citep{MSB08}:
\begin{displaymath}
S(CH) = -2.5 \log \frac{\int_{4280}^{4320} I_{\lambda}d\lambda}{\int_{4050}^{4100} I_{\lambda}d\lambda + \int_{4330}^{4350} I_{\lambda}d\lambda }
\end{displaymath}
which was defined for use with the low-metallicity globular cluster M53  ([Fe/H]=-1.84);

\noindent
$s_{CH}$ \citep{BS93}:
\begin{displaymath}
s_{CH} = -2.5 \log \frac{\int_{4280}^{4320} I_{\lambda}d\lambda}{\int_{4220}^{4280} I_{\lambda}d\lambda}
\end{displaymath}
which was defined for use with the moderate-metallicity globular cluster M13  ([Fe/H]=-1.54);

\noindent
$m_{CH}$ \citep{T83}:
\begin{displaymath}
m_{CH} = -2.5 \log \frac{(\frac{1}{50\mbox{\AA}}){\int_{4270}^{4320}F_{\nu} d\lambda}}{(\frac{1}{110\mbox{\AA}}){\int_{4020}^{4130}F_{\nu} d\lambda}+(\frac{1}{90\mbox{\AA}}){\int_{4430}^{4520}F_{\nu} d\lambda}}
\end{displaymath}
which was defined for use with red giants in metal-poor globular clusters M92 and M15 ([Fe/H] $\simeq -2.3$);

\noindent
$GP$ \citep{R05}:
\begin{displaymath}
GP = -2.5 \log \frac{\int_{4297.5}^{4312.5}I_{\lambda}}{\int_{4247}^{4267}I_{\lambda}d\lambda + \int_{4363}^{4372}I_{\lambda}d\lambda}
\end{displaymath}
defined for a study of carbon-enhanced metal-poor field stars;

\noindent
and $CH(4300)$ \citep{H03}:
\begin{displaymath}
CH(4300) = -2.5 \log \frac{\int_{4280}^{4320} I_{\lambda}d\lambda}{\int_{4250}^{4280} I_{\lambda}d\lambda + \int_{4320}^{4340} I_{\lambda}d\lambda }
\end{displaymath}
which was defined for use with main-sequence stars of the globular cluster 47 Tucanae. In the above equations, $I_{\lambda}$ is the flux per Angstrom, recorded in data such as \citet{MSB08} in units of ADU per pixel, and $F_{\nu}$ is the flux per unit frequency, calibrated to true energy units in the study of \citet{T83}. 

\section{Defining a new G-band index}
We are motivated to define a G band index that is valid across a wide range of metallicity for use in studies of carbon and nitrogen abundances in bright globular cluster red giants. Since this region of the spectrum has broad CN features as well as broad CH features, minimal dependence on nitrogen abundance is an important criterion for our index definition, in addition to predictable, monotonic dependence on overall stellar metallicity and [C/Fe]. This is doubly important in light of the work we intend to use this index for, converting CH and 3883$\mbox{\AA}$ CN bandstrengths to carbon and nitrogen abundances to study the rate of ongoing CN(O)-cycle processing in globular cluster red giants.

In the present analysis, we use a large grid of synthetic spectra generated with MARCS model atmospheres \citep{G75} and the SSG spectrum-synthesis code (Bell et al. 1994 \nocite{BPT94} and references therein). These are the same synthetic spectra used for index-based abundance analyses in \citet{MSB08}, \citet{CBS02}, and other papers by the current authors. Our particular grid represents bright red giants with $M_{V}=-1.5$, twelve values of [Fe/H] ranging from $-2.31$ to $-0.83$, [C/Fe] varying from $-1.4$ to $+0.4$ in steps of $0.2$ dex, and [N/Fe] ranging from $-0.6$ to $+2.0$, also in steps of $0.2$ dex. Other model parameters, identical for each model, are [O/Fe]$=+0.2$, [Ca/Fe]$=+0.3$, $^{12}{\rm C}/^{13}{\rm C}=10$, and $v_{\rm turb}=2$ km/s. Spectra were smoothed to a resolution of 5.4$\mbox{\AA}$ and a pixel spacing of 1.8$\mbox{\AA}$ to better match the data we intend to study with this index. Figure 1 shows the relevant region of two of the synthetic spectra, with [Fe/H]$=-1.41$, [C/Fe]$=+0.4$, and [N/Fe]$=-0.6$ (thin line) and $+2.0$ (heavy line), normalized at 4319.2$\mbox{\AA}$ to emphasize relative differences. The G band is marked by vertical dotted lines at the $S(CH)$ \citep{MSB08} science band, and the broad nitrogen-sensitive feature blueward of the G band is the 4215$\mbox{\AA}$ CN band. The upper panel of Figure 2 shows synthetic spectra with typical globular cluster values for [Fe/H] ($-1.41$ dex) and [C/Fe] ($-0.4$ dex), and the full range of [N/Fe] available in our model grid. The values of [N/Fe] range from $-0.6$ (weakest absorption) to $+2.0$ (strongest absorption) in steps of $0.2$ dex. The lower panel of Figure 2 shows an analogous set of spectra, with [N/Fe] fixed at $+0.6$ and [C/Fe] varying from $-1.4$ to $+0.4$ dex, the full range represented in our models, also in steps of 0.2 dex.  There is very little wavelength range that is unaffected by carbon or nitrogen in blue spectra of stars of this type.

\begin{figure}[]
\includegraphics[width=\columnwidth]{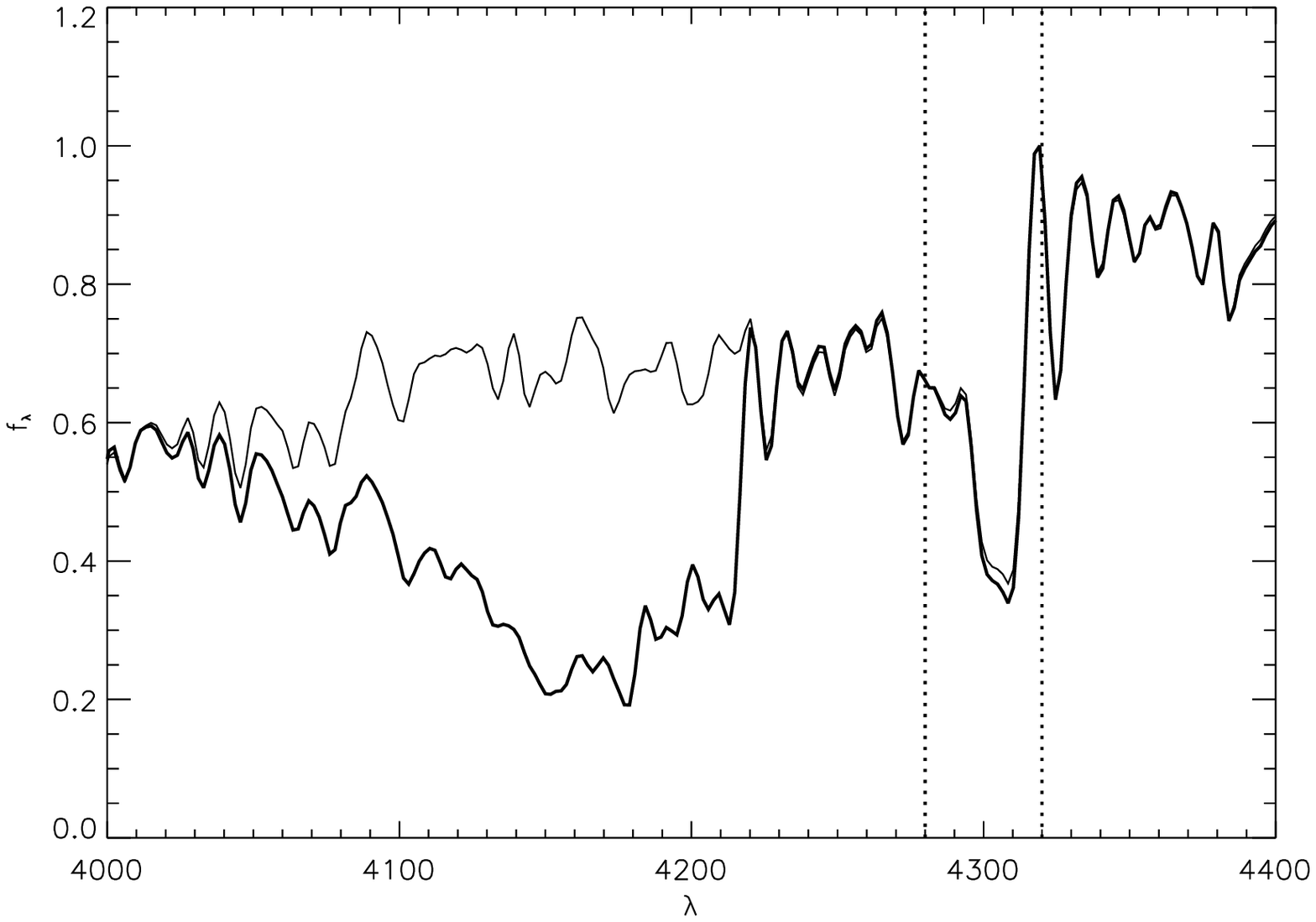}
\caption{The G-band region of two synthetic spectra from our set with [Fe/H]$=-1.41$, [C/Fe]$=+0.4$, and [N/Fe]$=-0.6$ (thin line) and $+2.0$ (heavy line). The vertical dotted lines mark the science band of the low-metallicity G-band index $S(CH)$ \citep{MSB08}. The broad nitrogen-sensitive feature to the blue of the G band is the 4215$\mbox{\AA}$ CN band. Spectra are normalized at 4319.2$\mbox{\AA}$, just redward of the G band, to emphasize nitrogen-sensitive features.
}
\end{figure}
\begin{figure}[]
\includegraphics[width=\columnwidth]{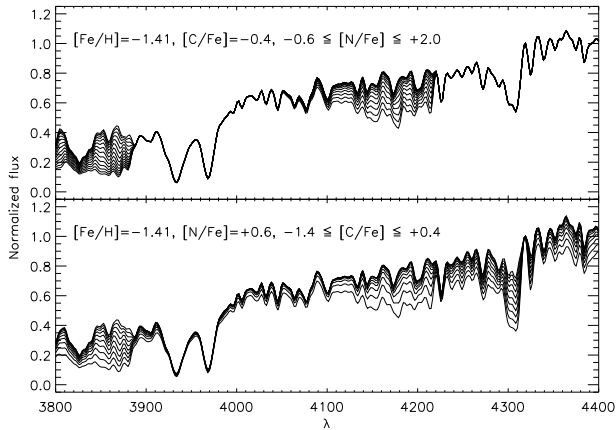} 
\caption{Synthetic spectra demonstrating the full range of [N/Fe] (upper panel) and [C/Fe] (lower panel) available in our model grid. Spectra are normalized just redward of the G band, as in Figure 1. Nitrogen sensitivity is almost entirely confined to the 3883$\mbox{\AA}$ and 4215$\mbox{\AA}$ CN bands, while carbon sensitivity is nearly universal.
}
\end{figure}

Figure 3 shows the CH index $S(CH)$ \citep{MSB08} versus nitrogen abundance for the full range of [C/Fe] and [N/Fe] available in our grid of synthetic spectra. In each panel, [C/Fe] increases from $-1.4$ (leftmost column) to $+0.4$ (rightmost column) in steps of 0.2 dex. The [Fe/H] metallicity varies from $-2.31$ (top left panel) to $-0.83$ (lower right), and as the overall metallicity increases, the resulting index-abundance grid becomes less rectangular. At the highest metallicity, high nitrogen abundance causes a large decrease in the $S(CH)$ index for a fixed carbon abundance, making it difficult to determine [C/Fe] from $S(CH)$ without first knowing [N/Fe]. This effect can be attributed to the presence of CN absorption features in the blue comparison bandpass of the $S(CH)$ index. Equivalent plots for the index $CH(4300)$ \citep{H03}, which was used for a study of main-sequence stars in 47 Tuc, are shown in Figure 4. The decrease in the measured index value seen in Figure 3 has been overcome(by placement of the blue comparison bandpass redward of the $\lambda$4215 CN bandhead), and slightly overcompensated for, producing a more regular grid of points for all but the largest values of [Fe/H], [C/Fe] and [N/Fe]. However, the horizontal range in these plots is smaller than in Figure 3, meaning that there is a smaller change in $CH(4300)$ for a given change in [C/Fe]. Ideally, our new index will combine the good characteristics of these two figures, with strong dependence on [C/Fe] and independence from [N/Fe].

\begin{figure}[]
\includegraphics[width=\columnwidth]{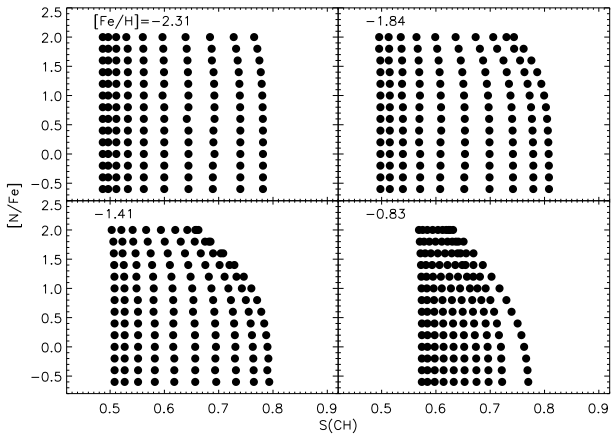} 
\caption{Low-metallicity G-band index $S(CH)$ \citep{MSB08} versus [N/Fe] for four example metallicities spanning our model grid. [C/Fe] ranges from $-1.4$ (leftmost column in each panel) to $+0.4$ (rightmost column in each panel) in steps of 0.2 dex. While $S(CH)$ is fairly unaffected by [N/Fe] at low metallicity, there is a marked decrease in $S(CH)$ at high values of [N/Fe] as [Fe/H] rises. This effect makes $S(CH)$ an unreliable indicator of [C/Fe] for stars in the high abundance range.
}
\end{figure}
\begin{figure}[]
\includegraphics[width=\columnwidth]{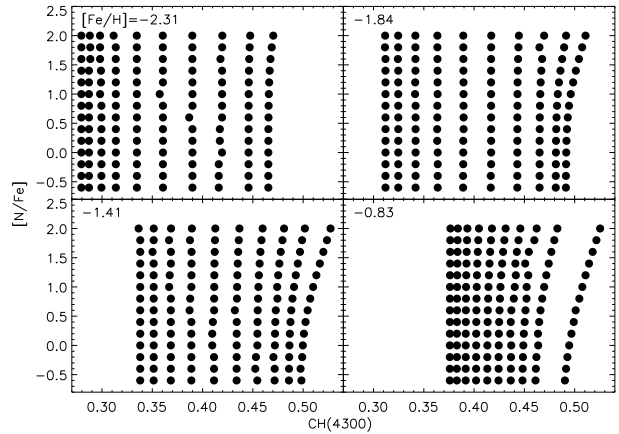} 
\caption{Main-sequence G-band index $CH(4300)$ \citep{H03} versus [N/Fe] for the same four example metallicities and [C/Fe] options shown in Figure 3. The effect of [N/Fe] on measured bandstrength is smaller than in Figure 3, but the sensitivity to [C/Fe] is also lower.
}
\end{figure}

\section{Optimization}
We use the general function minimizer TNMIN\footnote{Written and distributed by C. Markwardt; more curve-fitting and minimization routines available from http://astrog.physics.wisc.edu/\char126 craigm/idl/idl.html} to define a new index $S_{2}(CH)$ in a way that simultaneously maximizes its sensitivity to carbon abundance and minimizes the influence of the nearby CN features. The quantity that TNMIN uses to determine the nitrogen sensitivity of an index is $\frac{|\Delta S_{2}(CH)|}{\Delta {\rm [N/Fe]}}$, the mean of the absolute value of the change in $S_{2}(CH)$ per unit dex change in [N/Fe]. This parameter is calculated between consecutive [N/Fe] values in the subset of model grid points that have maximal [C/Fe] (e.g., $\frac{1}{0.2}(S_{2}(CH)|_{\rm [C/Fe]=+0.4, [N/Fe]=+2.0}-S_{2}(CH)|_{\rm [C/Fe]=+0.4, [N/Fe]=+1.8})$). As can be seen in Figure 3, the models with the largest [C/Fe] abundance (the rightmost column of points) are diverted the furthest from a rectangular grid by high [N/Fe] and [Fe/H] abundances. Minimizing the mean derivative between consecutive points in that column will force it to be as vertical as possible, which will mean that the definition for $S_{2}(CH)$ will be as nitrogen-independent as possible. A similar quantity $\frac{\Delta S_{2}(CH)}{\Delta {\rm [C/Fe]}}$ is our measure of the sensitivity of $S_{2}(CH)$ to carbon abundance. It defined to be the difference between $S_{2}(CH)$ measured at maximal [C/Fe] and minimal [N/Fe], and $S_{2}(CH)$ measured at minimal [C/Fe] and minimal [N/Fe], divided by the range of [C/Fe] available in our model grid. In Figure 3, for example, $\frac{\Delta S_{2}(CH)}{\Delta {\rm [C/Fe]}}$ is measured as the difference in $S_{2}(CH)$ between the rightmost and leftmost points of the lowest row, divided by 1.8 dex. A large value for $\frac{\Delta S_{2}(CH)}{\Delta {\rm [C/Fe]}}$ will correspond to a $S_{2}(CH)$-[N/Fe] plot covering a large horizontal range, and will reduce the error inherent in using $S_{2}(CH)$ to determine [C/Fe]. 

We assume a priori that minimizing the effect of [N/Fe] on CH bandstrength will mean keeping the comparison and science bands from overlapping with  the CN features, primarily the 4215$\mbox{\AA}$ band, while maximizing the response to carbon will mean restricting the science band to the G band. We approached the two CH bandstrength criteria independently, first fixing the location of the science band and allowing TNMIN to find the best locations for the comparison bands by minimizing nitrogen dependence, and then fixing the comparison bands and having TNMIN return the optimal science band location by maximizing carbon dependence. This process is then repeated, with the results of the previous optimization used as inputs for the next, until the result is self-consistent. The problem is solved independently for each of the twelve [Fe/H] values in the synthetic spectrum grid, and we find that the result is not strongly dependent on overall stellar metallicity.

\begin{figure}[]
\includegraphics[width=\columnwidth]{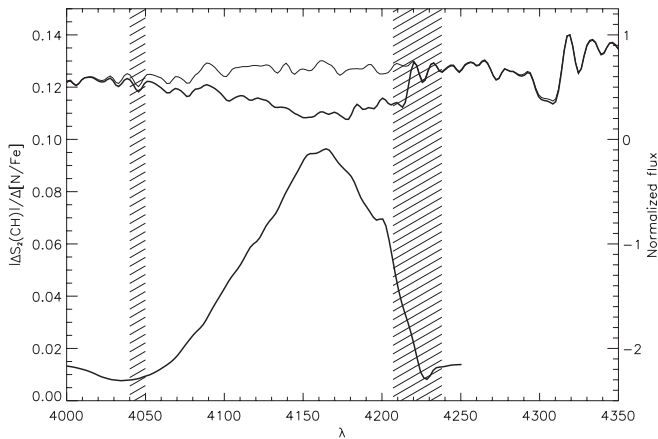} 
\caption{Nitrogen sensitivity measure $\frac{|\Delta S_{2}(CH)|}{\Delta {\rm [N/Fe]}}$ versus central wavelength for a 30-$\mbox{\AA}$-wide blue comparison band, calculated assuming a science band from 4297$\mbox{\AA}$ to 4317$\mbox{\AA}$ and a red comparison band from 4330$\mbox{\AA}$ to 4375$\mbox{\AA}$, for the typical globular cluster metallicity [Fe/H]$=-1.41$. Shaded regions show the two blue comparison band locations preferred by the TNMIN routine. Synthetic spectra with [Fe/H]$=-1.41$, [C/Fe]$=+0.4$, and [N/Fe]$=-0.6$ (thin line) and $+2.0$ (heavy line) are also plotted (uppermost two curves) to show the spectral behavior inside the possible comparison bands.
}
\end{figure}

There are two virtually equivalent options for the location of the blue comparison band returned by the TNMIN program, 4040$\mbox{\AA}$ to 4050$\mbox{\AA}$ and 4207$\mbox{\AA}$ to 4238$\mbox{\AA}$. Since the index we are trying to define will be used on data with a signal to noise ratio per pixel between 40 and 100, we would like for the bands to be wide, so that pixel-to-pixel random errors will be averaged out. We would also like for the bands to be closely-spaced, so that errors in flux calibration will be minimally important, making the resulting bandstrength measurement more easily intercomparable with data taken on other dates, with other instruments and by other observers. By both of these criteria, placing the blue sideband at 4207$\mbox{\AA}$ to 4238$\mbox{\AA}$ will produce a better index. To further explore the dependence of $\frac{|\Delta S_{2}(CH)|}{\Delta {\rm [N/Fe]}}$ on the placement of the blue sideband, we calculated $\frac{|\Delta S_{2}(CH)|}{\Delta {\rm [N/Fe]}}$ for all twelve [Fe/H] values available in our model grid, using 151 definitions of $S_{2}(CH)$ with a red comparison band at 4330$\mbox{\AA}$ to 4375 $\mbox{\AA}$, the $S(CH)$ science band (4280$\mbox{\AA}$ to 4320$\mbox{\AA}$), and a 30-$\mbox{\AA}$-wide blue comparison band centered at wavelengths ranging from 3950.5$\mbox{\AA}$ to 4250.5$\mbox{\AA}$. This produced twelve $\frac{|\Delta S_{2}(CH)|}{\Delta {\rm [N/Fe]}}$ versus $\lambda_{center}$ curves, which are averaged together to create the curve shown in Figure 5, which also includes TNMIN's preferred locations for the blue comparison band (shaded regions) as well as synthetic spectra with [Fe/H]$=-1.4$, [C/Fe]$=+0.4$, and [N/Fe]$=-0.6$ (thin line) and $+2.0$ (heavy line). The two relative minima in the $\frac{|\Delta S_{2}(CH)|}{\Delta {\rm [N/Fe]}}$ curve correspond relatively well with the TNMIN results, and with expectations: both are at the edges of the 4215$\mbox{\AA}$ CN band.

\begin{figure}[]
\includegraphics[width=\columnwidth]{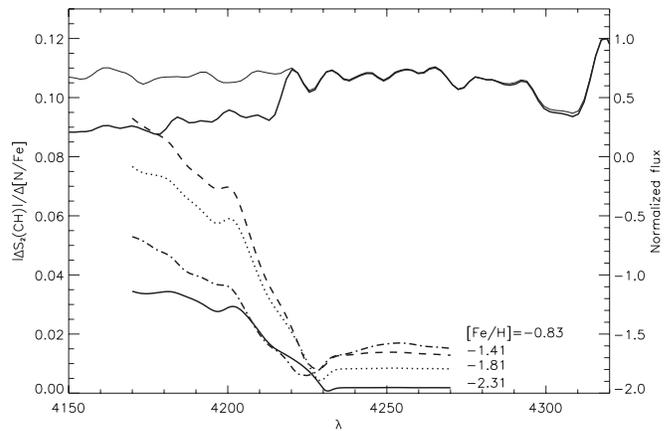} 
\caption{More detailed version of Figure 5, with $\frac{|\Delta S_{2}(CH)|}{\Delta {\rm [N/Fe]}}$ versus central wavelength for a 30-$\mbox{\AA}$-wide comparison band, calculated for four sample metallicities spanning the range of our model grid. The location of the minimum in each curve (which produces the least nitrogen-sensitive G-band index) is dependent on [Fe/H]. 
}
\end{figure}

Figure 6 shows a subset of our models in more detail: the four lower curves are $\frac{|\Delta S_{2}(CH)|}{\Delta {\rm [N/Fe]}}$ calculated for a 30-$\mbox{\AA}$-wide band centered at a wavelength varying from 4170$\mbox{\AA}$ to 4270$\mbox{\AA}$, for the same four metallicities used in Figure 3. The ideal location, the central wavelength where $\frac{|\Delta S_{2}(CH)|}{\Delta {\rm [N/Fe]}}$ is minimal, is dependent on [Fe/H], with higher-metallicity curves preferring a comparison band that overlaps more with the 4215$\mbox{\AA}$ CN band. In addition, the minimum value of $\frac{|\Delta S_{2}(CH)|}{\Delta {\rm [N/Fe]}}$ is also dependent on metallicity: the lower minima of the lower-metallicity curves mean decreases with decreasing metallicity, meaning that, as seen in Figures 3 and 4, $S_{2}(CH)$ is less dependent on [N/Fe] at low metallicity. Interestingly, optimal comparison band locations at all metallicities include a small region of the 4215$\mbox{\AA}$ CN bandhead in the blue comparison band, and all of the $\frac{|\Delta S_{2}(CH)|}{\Delta {\rm [N/Fe]}}$ curves level off at non-minimal values as soon as the blue comparison band moves entirely out of the 4215$\mbox{\AA}$ absorption band. This is unexpected given that we are explicitly designing $S_{2}(CH)$ to be independent of nitrogen abundance. The implication is that either the science band or the red comparison band contains a region of spectrum that is sensitive to [N/Fe], and it is necessary to have a slightly nitrogen-sensitive blue comparison band to compensate. To choose a location for the blue comparison band that will work equally well for all metallicities, we once again average the $\frac{|\Delta S_{2}(CH)|}{\Delta {\rm [N/Fe]}}$ curves calculated for all twelve [Fe/H] values together. That mean curve is minimized when the blue comparison band runs from 4211.5$\mbox{\AA}$ to 4241.5$\mbox{\AA}$. This is close to, but not exactly the same as, the result of our recursive TNMIN process. The difference comes from the use of metallicity-averaged $\frac{|\Delta S_{2}(CH)|}{\Delta {\rm [N/Fe]}}$ curves in this subsequent refinement of the TNMIN result.

The location of the red comparison band returned by TNMIN is virtually the same as in the definitions of \citet{H03} and \citet{MSB08}: 4330$\mbox{\AA}$ to 4375$\mbox{\AA}$. Since there is very little [C/Fe]-independent spectrum redward of the G band (see Figure 2), we adopt this result directly as the optimal location for the red comparison band instead of trying to improve on it.

\begin{figure}[]
\includegraphics[width=\columnwidth]{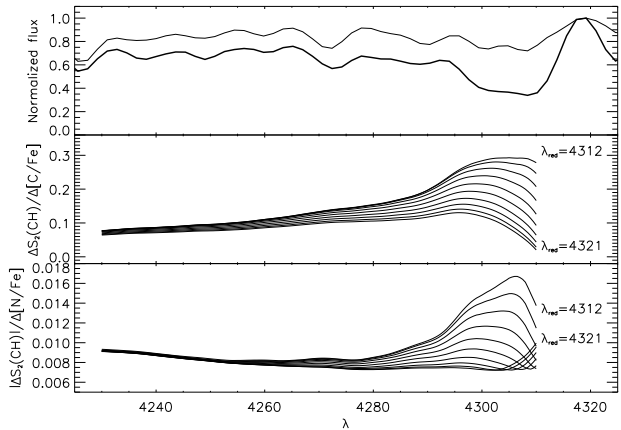} 
\caption{Metallicity-averaged $\frac{\Delta S_{2}(CH)}{\Delta {\rm [C/Fe]}}$ (middle panel) and $\frac{|\Delta S_{2}(CH)|}{\Delta {\rm [N/Fe]}}$ (lower panel) curves versus wavelength $\lambda_{blue}$ of the blue end of the science bandpass. The top panel plots synthetic spectra for [Fe/H]$=-1.41$, [N/Fe]$=+2.0$, and [C/Fe]$=-1.4$ (thin line) and $+0.4$ (heavy line) overplotted to show the location of the G band. Each individual curve is calculated by fixing the red edge of the $S_{2}(CH)$ science band at a value between 4312$\mbox{\AA}$ and 4321$\mbox{\AA}$, then varying the blue edge from 4230$\mbox{\AA}$ to 4310$\mbox{\AA}$. Comparison bands are held fixed at their optimal locations (4212$\mbox{\AA}$ to 4242$\mbox{\AA}$ and 4330$\mbox{\AA}$ to 4375$\mbox{\AA}$). As described in the text, the red edge of the science band is set based on the $\frac{|\Delta S_{2}(CH)|}{\Delta {\rm [N/Fe]}}$ curves, and the blue edge location is based on the $\frac{\Delta S_{2}(CH)}{\Delta {\rm [C/Fe]}}$ curves.
}
\end{figure}

Fixing the comparison bands in their optimal locations and allowing the science band location to vary, we find that TNMIN strongly prefers a narrow science band. This is reasonable; the strongest part of the CH feature has the strongest dependence on [C/Fe]. However, it is impractical for the science band to be too narrow; as with the comparison bands, we want to have a broad enough wavelength coverage that noise in the spectrum will average out to zero. To better understand the optimal placement of the $S_{2}(CH)$ science band, we calculated the carbon- and nitrogen-sensitivity measures $\frac{\Delta S_{2}(CH)}{\Delta {\rm [C/Fe]}}$ and $\frac{|\Delta S_{2}(CH)|}{\Delta {\rm [N/Fe]}}$ for a wide variety of possible $S_{2}(CH)$ index definitions. In each case, the comparison bands were fixed at their optimal locations, as discussed above. For ten values of $\lambda_{red}$ (the position of the red edge of the science band) between 4312$\mbox{\AA}$ and 4321$\mbox{\AA}$, we calculated $\frac{\Delta S_{2}(CH)}{\Delta {\rm [C/Fe]}}$ and $\frac{|\Delta S_{2}(CH)|}{\Delta {\rm [N/Fe]}}$ for 161 possible values of $\lambda_{blue}$ (the position of the left edge of the science band) and all twelve [Fe/H] values available in our model grid. We then collapsed the twelve independent [Fe/H] curves corresponding to each value of $\lambda_{red}$ into ten metallicity-averaged curves. We base our determination of the optimal values for $\lambda_{red}$ and $\lambda_{blue}$ on these mean curves.

\begin{figure}[]
\includegraphics[width=\columnwidth]{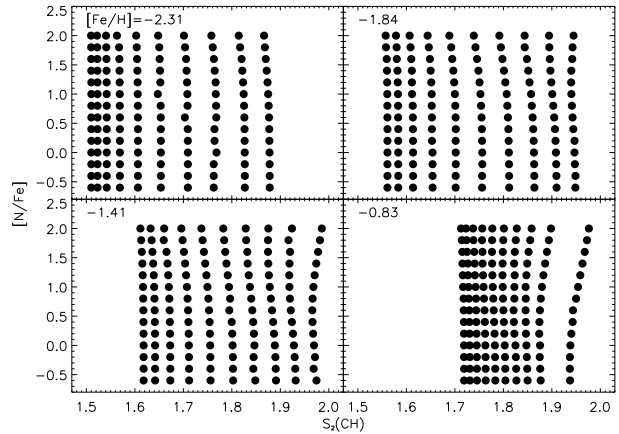} 
\caption{In analogy to Figures 3 and 4, this plot shows $S_{2}(CH)$ versus [N/Fe] for four sample metallicities and ten [C/Fe] values spanning our model grid. The decrements in index value seen at high [Fe/H] and [N/Fe] in Figure 3 have been reduced at moderate [Fe/H] and replaced by a slight increase at high [Fe/H], and the horizontal range $\frac{\Delta S_{2}(CH)}{\Delta {\rm [C/Fe]}}$ is larger at all metallicities than in Figures 3 and 4.
}
\end{figure}

The ten metallicity-averaged sensitivity curves are shown in the middle and lower panels of Figure 7 as a function of $\lambda_{blue}$. In the upper panel synthetic spectra are shown as a function of $\lambda$ for [Fe/H]$=-1.41$, [N/Fe]$=+2.0$, and [C/Fe]$=-1.4$ (thin line) and $+0.4$ (heavy line). Generally, lower values of $\lambda_{red}$ result in larger values for both $\frac{\Delta S_{2}(CH)}{\Delta {\rm [C/Fe]}}$ and $\frac{|\Delta S_{2}(CH)|}{\Delta {\rm [N/Fe]}}$, though the $\frac{|\Delta S_{2}(CH)|}{\Delta {\rm [N/Fe]}}$ curves with larger $\lambda_{red}$ values show an upturn for values of $\lambda_{blue}$ above 4300$\mbox{\AA}$. The rise in $\frac{\Delta S_{2}(CH)}{\Delta {\rm [C/Fe]}}$ as the blue edge of the science band approaches the strongest part of the G band is a clear visual representation of TNMIN's preference for a narrow science band for $S_{2}(CH)$.

Our criteria for the science band, in order of importance, are a low value for $\frac{|\Delta S_{2}(CH)|}{\Delta {\rm [N/Fe]}}$, a high value for $\frac{\Delta S_{2}(CH)}{\Delta {\rm [C/Fe]}}$, and an overall bandwidth larger than 15$\mbox{\AA}$. We select $\lambda_{red}$ by requiring that $\frac{|\Delta S_{2}(CH)|}{\Delta {\rm [N/Fe]}}$ be lower than 0.01 in the range 4290$\mbox{\AA}$$\leq \lambda_{blue} \leq$ 4300$\mbox{\AA}$, then choosing the smallest value of $\lambda_{red}$ that qualifies to allow for the largest possible $\frac{\Delta S_{2}(CH)}{\Delta {\rm [C/Fe]}}$. This results in a value for $\lambda_{red}$ of 4318$\mbox{\AA}$. To choose $\lambda_{blue}$, we simply find the highest point on the $\frac{\Delta S_{2}(CH)}{\Delta {\rm [C/Fe]}}$ curve corresponding to $\lambda_{red}=4317$$\mbox{\AA}$, which gives a value for $\lambda_{blue}$ of 4297$\mbox{\AA}$. This result satisfies our bandstrength criterion. We then performed a check for self-consistency, using this final science band location as the input for placement of the blue comparison band. We found that the resulting comparison band was nearly identical to our original result (4212$\mbox{\AA}$ to 4242$\mbox{\AA}$). Repeating the process of fixing the comparison bands to choose a science band, we find that it also shifts slightly. We adopt the range 4297$\mbox{\AA}$ to 4317$\mbox{\AA}$ as the optimal location for the science band of $S_{2}(CH)$, producing a final bandstrength definition of

\begin{displaymath}
S_{2}(CH) = -2.5 \log \frac{\int_{4297}^{4317} I_{\lambda}d\lambda}{\int_{4212}^{4242} I_{\lambda}d\lambda + \int_{4330}^{4375} I_{\lambda}d\lambda }
\end{displaymath}

Figure 8 is analogous to Figures 3 and 4, and shows $S_{2}(CH)$ versus [N/Fe] for the full [C/Fe] and [N/Fe] range available in our grid of synthetic spectra, at four example metallicities. The points in this plot, in particular those at high [C/Fe] and high [N/Fe], are arranged much more uniformly and perpendicularly than those shown in Figures 3 and 4. The horizontal range covered is also larger than in Figures 3 and 4, meaning that $S_{2}(CH)$ is more sensitive to [C/Fe] than $S(CH)$ or $CH(4300)$. As a result, the transformation from measured $S_{2}(CH)$ to [C/Fe] is more straightforward and less error-prone.

The lower right panel of Figure 8, showing $S_{2}(CH)$ versus [N/Fe] for synthetic spectra with [Fe/H]$=-0.83$, does show deviation from rectangularity at large [C/Fe] and [N/Fe] values. Fortunately, in the stars this index is designed for, this abundance combination is rarely, if ever, observed to exist. In bright red giants in globular clusters, large nitrogen abundances are strongly correlated with low carbon abundances (in general, [C+N+O/Fe] is roughly constant in these stars, e.g., Denissenkov et al. 1998\nocite{D98} and Sneden et al. 1997\nocite{S97}), though the mechanism for producing these particular abundance ratios is a subject of ongoing study. As a result, this unavoidable imperfection in our index definition should not present a large problem for studies of bright red giants in globular clusters.

\begin{figure}[]
\includegraphics[width=\columnwidth]{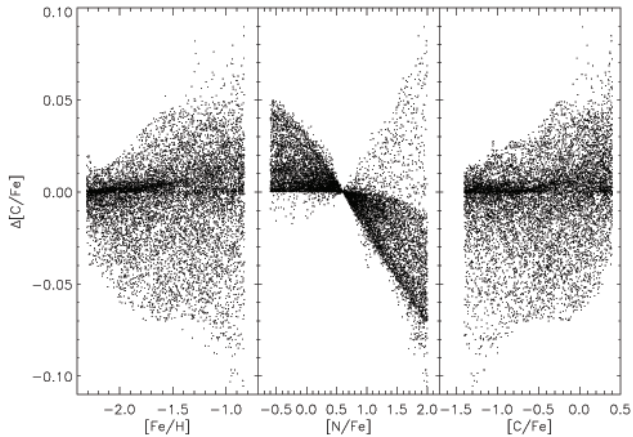} 
\caption{Method-based measurement error $\Delta$[C/Fe] versus [Fe/H], [C/Fe], and [N/Fe] for $10^{4}$ randomly-chosen ([Fe/H], [C/Fe], [N/Fe]) abundance points within the range of our model grid. The error $\Delta$[C/Fe] is a complex function of abundance, but is never prohibitively large: 99.7\% of points with typical globular cluster red giant abundances ($-2.31 \leq$ [Fe/H] $\leq -0.83$, $-1.0 \leq$ [C/Fe] $\leq 0.0$, $0.0 \leq$ [N/Fe] $\leq +1.0$) fall in the range $-0.03 \leq \Delta$[C/Fe] $\leq +0.03$.
}
\end{figure}

\section{Applications and limitations of this index definition}
In order to characterize the error introduced by using $S_{2}(CH)$ with an assumed nitrogen abundance to determine [C/Fe], we performed a series of Monte Carlo tests, in each trial selecting $10^{4}$ random ([Fe/H], [N/Fe], [C/Fe]) abundance sets within the parameter space occupied by our synthetic spectra and then interpolating within the model grid to find the corresponding ''observed'' values of $S_{2}(CH)$. We used a process similar to the one described in \citet{MSB08} to convert each randomly-chosen value of $S_{2}(CH)$ to [C/Fe], first interpolating within the model grid to match the randomly-chosen [Fe/H], then assuming a [N/Fe] value of $+0.6$ and interpolating linearly within that subgrid to find the [C/Fe] value needed to produce the ''observed'' $S_{2}(CH)$. The difference between the calculated and original input [C/Fe] values is the error $\Delta$[C/Fe]. We also performed this test with the extra step of adding Poisson-distributed 1\%, 2\%, 5\% and 10\% noise to the synthetic spectrum corresponding to the randomly-chosen abundance set and redetermining the value of $S_{2}(CH)$, to understand how much additional error in the calculated [C/Fe] might be introduced by noise in the data.

\begin{figure}[]
\includegraphics[width=\columnwidth]{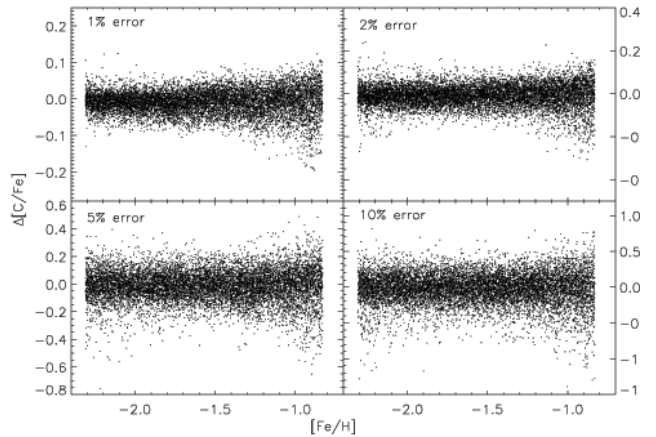} 
\caption{Each panel shows $\Delta$[C/Fe] versus [Fe/H] for $10^{4}$ randomly-chosen ([Fe/H], [C/Fe], [N/Fe]) abundance points within our model grid. We added Poisson noise to the spectra corresponding to those abundances, then calculated [C/Fe] and $\Delta$[C/Fe] based on the value of $S_{2}(CH)$ measured from the noisy spectra. The noise ranges from 1\% (upper left) to 10\% (lower right). While the shape of the $\Delta$[C/Fe] distribution remains the same in all panels, the magnitude is strongly dependent on the amount of noise added to the spectrum. The standard deviation $\sigma_{\Delta{\rm [C/Fe]}}$ roughly doubles in each successive panel. 
}
\end{figure}

Figure 9 shows $\Delta$[C/Fe] versus [Fe/H], [N/Fe], and [C/Fe] for the no-noise test, and there are two things immediately obvious from these figures. First, that the behavior of $\Delta$[C/Fe] is complex and not simply related to any one parameter. Second, $S_{2}(CH)$ is fairly reliably converted to [C/Fe]: the absolute value of $\Delta$[C/Fe] is never larger than 0.11 dex, and the vast majority of errors are quite small. The mean $\Delta$[C/Fe] value is $-0.0049$, and 80\% of all trials were in the range $-0.03 \leq \Delta$[C/Fe] $\leq 0.03$. 

Those numbers change to $-0.024$ and 58\% in trials with [N/Fe]$\geq 1.0$, to $-0.0031$ and 72\% in trials with [Fe/H]$\geq -1.4$, and to $0.0046$ and 78\% in trials with [C/Fe] $\geq 0.0$. In the region of abundance space typically occupied by globular cluster red giants ($-2.31 \leq$ [Fe/H] $\leq -0.83$, $-1.0 \leq$ [C/Fe] $\leq 0.0$, $0.0 \leq$ [N/Fe] $\leq 1.0$ ), the mean $\Delta$[C/Fe] value is $0.0020$, and 99.7\% of trials have $-0.03 \leq \Delta$[C/Fe] $\leq 0.03$.

Figure 10 shows $\Delta$[C/Fe] as a function of [Fe/H] for one particular Monte Carlo realization of each Poisson noise-added test. In tests with 1\% and 2\% noise added to the spectra, the $S_{2}(CH)$-to-[C/Fe] conversion process is still quite reliable. Mean values for $\Delta$[C/Fe] are $-0.0089$ and $0.0098$, respectively, and 66\% and 49\% of points, respectively, fall in the range $-0.03 \leq \Delta$[C/Fe] $\leq 0.03$. Considering only the region of abundance space likely to be occupied by globular cluster red giants, those numbers change to $-0.0008$ and 84\% (for 1\% error) and $-0.0004$ and 60\% (for 2\% error). The tests with 5\% and 10\% error are less encouraging: while the shape of the $\Delta$[C/Fe] distribution remains the same in all panels of Figure 10, the standard deviation $\sigma_{\Delta{\rm [C/Fe]}}$ roughly doubles with each increase in noise. Again restricting the abundance range to the region where globular cluster red giants are mostly likely to be found, the mean value for $\Delta$[C/Fe] and percentage of stars falling in the range $-0.03 \leq \Delta$[C/Fe] $\leq 0.03$ are $-0.0067$ and 27\%, respectively, for the 5\%-error case and $-0.0067$ and 14\%, for the 10\%-error case. This implies that [C/Fe] abundances derived from $S_{2}(CH)$ (or any other G-band index) should only be trusted when the original spectra have decently high signal to noise ratios: $\Delta$[C/Fe] is unacceptably large in the 10\% error case, and is marginal in stars with typical globular cluster abundances in the 5\% error case.

\begin{figure}[]
\includegraphics[width=\columnwidth]{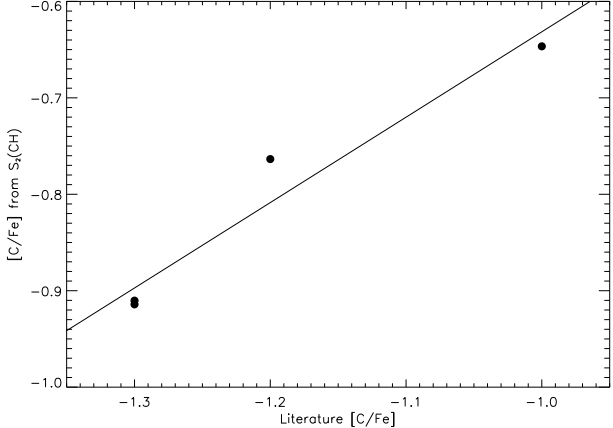} 
\caption{A comparison of [C/Fe] values determined using $S_{2}(CH)$, and [C/Fe] values from \citet{SB06}, for four giants in M13 observed by the authors as part of an ongoing study. While there is a fairly predictable relationship between the two sets of measurements, there is clearly an offset between the zeropoints of the two carbon abundance scales.
}
\end{figure}

It is important to test our new index on data, and not just noise-injected synthetic spectra, to demonstrate that it does return reliable [C/Fe] abundance values. Figure 11 shows a comparison of [C/Fe] values derived from $S_{2}(CH)$ with [C/Fe] values taken from the literature compilation of \citet{SB06} (originally reported in Suntzeff 1981\nocite{S81}), for four M13 giants observed by the present authors as part of an ongoing study of carbon depletion in globular cluster red giants. The spectra were taken with the Kast double spectrograph on the 3-meter Shane telescope at Lick Observatory, and have moderate resolution (R$\simeq 1000$). The synthetic spectra were smoothed to match the data resolution and pixel spacing. Carbon abundances were derived from measured $S_{2}(CH)$ bandstrengths through an index-matching process similar to the one described in \citet{MSB08}.

Figure 11 shows a clear linear relationship between the \citet{S81} [C/Fe] abundances and the [C/Fe] values based on $S_{2}(CH)$ for the four stars in common between the samples, though the abundance scales defined by the two different abundance-determination methods have different zeropoints. The magnitude of the offset between the two abundance scales is quite similar to that reported in \citet{SB06}, roughly $-0.35$ dex. In \citet{SB06} the authors find that abundance-scale offset between \citet{S81} [C/Fe] abundances and those reported in studies using MARCS model atmospheres \citep{G75}. Our carbon abundances based on $S_{2}(CH)$ can therefore be readily compared with other MARCS-based abundance determinations, and reliably transformed onto the abundance scale of \citet{S81} and compared with other measurements made on that scale.

Figure 12 shows method-based errors for carbon abundances as a function of [Fe/H], analogous to the leftmost panel of Figure 9, but expanded to show the same result calculated using the indices $S(CH)$ and $s_{CH}$. Both of these indices were defined for studies of one particular globular cluster, $S(CH)$ for the low-metallicity cluster M53 and $s_{CH}$ for the moderate-metallicity cluster M13, and therefore return small errors at and below those clusters' respective metallicities. The leftmost panel in Figure 12 is identical to the leftmost panel in Figure 9, and the vertical range is the same in all three panels. The mean value in the central panel of Figure 12, the difference between input and measured carbon abundances calculated using $S(CH)$, is -0.035, and 69\% of trials fall in the range $-0.03 \leq \Delta$[C/Fe]$_{1} \leq 0.03$. In the rightmost panel, which shows the difference between input and measured carbon abundances calculated from $s_{CH}$, the mean value is 0.007, and 86\% of trials have $-0.03 \leq \Delta$[C/Fe]$_{2} \leq 0.03$. The advantage of the $S_{2}(CH)$ index over these other two is that it has been designed to be fairly insensitive to [Fe/H] and [N/Fe], resulting in a smaller overall range of $\Delta$[C/Fe] errors across a wide range of [Fe/H], [N/Fe] and [C/Fe].

\begin{figure}[]
\includegraphics[width=\columnwidth]{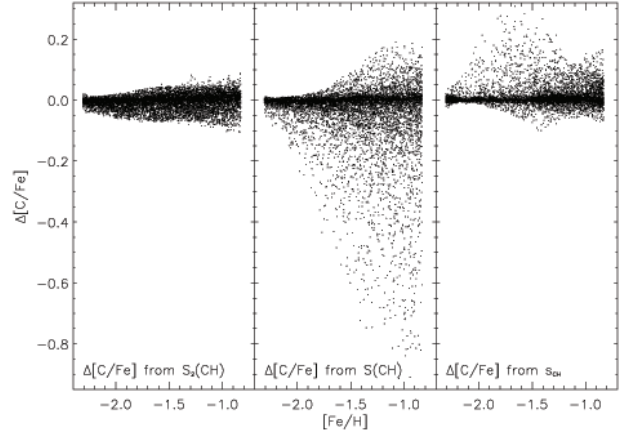} 
\caption{Each panel shows $\Delta$[C/Fe] versus [Fe/H] for $10^{4}$ randomly-chosen ([Fe/H], [C/Fe], [N/Fe]) abundance points within our model grid. In the leftmost panel, $\Delta$[C/Fe] was calculated from measurements of $S_{2}(CH)$; in the center, from $S(CH)$, and on the right, $s_{CH}$. The vertical scale in all three panels is identical.
}
\end{figure}

\section{Conclusions}
It is our hope that this new G-band index can provide a standard tool for low-resolution spectroscopic studies of bright red giants in globular clusters. It is simultaneously more sensitive to [C/Fe] and less sensitive to [N/Fe] than similar indices used in the literature to date, and would allow for useful comparison and combination of independent researchers' data in multiple globular clusters. The optimization method used here could be generally applicable to any spectral region where there are a few main parameters controlling the strength of an absorption or emission feature. Our characterizations of the dependence of $S_{2}(CH)$ on [Fe/H], [C/Fe], and [N/Fe], as well as the error associated with converting $S_{2}(CH)$ to [C/Fe], apply over a wide abundance range, but they are only strictly true for stars with $M_{V}=-1.5$, since the model atmospheres and synthetic spectra were only calculated for that particular value. While our current model calibration of $S_{2}(CH)$ should lead to a reliable measure of [C/Fe] for a moderate range in $M_{V}$, perhaps $-1.0 \geq M_{V} \geq -2.0$, it should not be applied to stars outside this magnitude range without recalibration. Since the index was designed specifically for bright red giants of Population II metallicities, we do not recommend using it to study [C/Fe] in stars belonging to significantly different populations.

\end{document}